\documentclass[nofootinbib,longbibliography,reprint,aps,prd,showpacs,superscriptaddress,groupedaddress]{revtex4-1}
\usepackage{amssymb,mathrsfs,amsmath}
\usepackage{braket}
\usepackage{graphicx}
\usepackage{dcolumn}   
\usepackage{bm} 
\usepackage{multirow}
\usepackage{footnote}
\usepackage{epstopdf}
\usepackage{rotating}
\usepackage{hyperref}
\usepackage{url}
\usepackage{xcolor}
\usepackage{pifont}
\usepackage[T1]{fontenc}
\usepackage{times}      

\graphicspath{ {./images/} }

\definecolor{darkblue}{cmyk}{1, 1, 0, 0}
\hypersetup{colorlinks=true,urlcolor=darkblue,citecolor=darkblue,linkcolor=darkblue}

\newcommand{\apjl}{Astrophys.~J.~Lett.}
\newcommand{\mnras}{Mon.~Not.~R.~Astron.~Soc.}
\newcommand{\aap}{Astron.~Astrophys.}

\newcommand{\physrep}{Phys.~Rep.}

\newcommand{\ssr}{Space Science Reviews}
\newcommand{\jcap}{J.~Cosmol.~Astropart.~Phys.}
\newcommand{\rmxaa}{Rev.~Mexicana~Astron.~Astrofis.}
\newcommand{\pasp}{Publ.~Astron.~Soc.~Pac.}

\newcommand{\nsigma}{\mathop{\mbox{$n$-$\sigma$}}}

\begin{document}
\title{Cosmological discordances II: Hubble constant, Planck and large-scale-structure data sets}
\date{\today}
\author{Weikang Lin}
\email{wxl123830@utdallas.edu}
\author{Mustapha Ishak}
\email{mishak@utdallas.edu}
\affiliation{Department of Physics, The University of Texas at Dallas, Richardson, Texas 75080, USA}

\begin{abstract}
We examine systematically the (in)consistency between cosmological constraints as obtained from various current data sets of the expansion history, large-scale structure (LSS), and cosmic microwave background (CMB) temperature and polarization from Planck. We run (dis)concordance tests within each set and across the three sets using a recently introduced index of inconsistency (IOI) capable of dissecting inconsistencies between two or more data sets. First, we compare the constraints on $H_0$ from five different methods and find that the IOI drops from 2.85 to 0.88 (on Jeffreys's scales) when the local $H_0$ measurements is removed. This seems to indicate that the local measurement is an outlier compared to the others, thus favoring a systematics-based explanation. We find a moderate inconsistency (IOI=2.61) between Planck temperature and polarization data sets. We find that current LSS data sets including the WiggleZ power spectrum, SDSS redshift space distortion, CFHTLenS weak lensing, CMB lensing, and cluster count from SZ effect, are consistent one with another and also when all combined. However, we find a persistent moderate inconsistency between Planck and individual or combined LSS probes. For Planck TT+lowTEB versus individual LSS probes, the IOI spans the range 2.92--3.72 and increases to 3.44--4.20 when the polarization data is added in. The joint LSS versus the combined Planck temperature and polarization has an IOI of 2.83 in the most conservative case. But if Planck low-$\ell$ temperature and polarization is also added to the joint LSS to constrain $\tau$ and break degeneracies, the inconsistency between Planck and joint LSS data increases to the high end of the moderate range with IOI=4.81.
Whether due to systematic effects in the data or to the underlying model, these inconsistencies need to be resolved. Finally, we perform forecast calculations using the Large Sky Synoptic Survey (LSST) and find that the discordance between Planck and future LSS data, if it persists as present, can rise up to a high IOI of 17, thus falling in the strong range of inconsistency.  

\end{abstract}

\maketitle

\section{Introduction}\label{section-IOI2-introduction}

Astronomical surveys and missions are getting more technologically advanced and sophisticated providing us with a plethora of observations and data sets. This allows one to require not only highly precise cosmological constraints but also highly consistent and accurate ones. In addition to parameter constraint analyses, we are witnessing an interest from the scientific literature in consistency tests between various data sets \cite{2006-Ishak-splitting, 2006-Marshall-etal-Bayesian,2011Robustness-March-etal, 2013-tension-Verde-etal,2015-discordance-MacCrann-etal,2014-rel-entropy-Seehars-etal,2016-quantify-concor-Seehars-etal,2016-Grandis-information-Gains,2016-tensions-Grandis-etal,2017MNRAS-Joudaki-etal-DIC-inconsistency,2017-Lin-Ishak-IOI-A,2007-Wang-etal-consistency,2015Ruiz-etal-param-splitting,2016Bernal-etal-param-splitting,2012-Shafieloo-crossing-stat,2016-Shafieloo-Hazra-Consistency-Planck,2017-Anand-Chaubal-Mazumdar-Mohanty}. Consistency tests have the potential to identify and localize systematic effects within particular data sets or to signal any issue with the underlying model; see, for example, a discussion in \cite{2016-Pogosian-Silvestri-consistency-Horndeski,DossettEtAlFOM,2016-leauthaud-etal-lensing-low,2016-Hubble-reconcile,2015bao-sdssIII,2016-Valentino-Melchiorri-Silk-reconciling-Planck-local,2017-Lopez-Corredoira-test-of-SMcosmo,2006-Ishak-splitting,2013Amendola-Marra-Quartin-Internal-robustness,2017-Lin-Ishak-IOI-A,2013-tension-Verde-etal,2015-rev-Joyce-Jain,DossettEtAlCFHTLenS,2016-Joyce-Lombriser-Schmidt-DEvsMG,DossettEtAlDEP,2016-Debono-Smoot-GRandCosmology,2007-Wang-etal-consistency,2015Ruiz-etal-param-splitting,DossettEtAlCurvature,2015-Valentino-Melchorri-Silk-Beyon-LCDM,Interacting.DE.DM2,2017-Zhao-etal-nature-astronomy,2017-Khosravi-etal-H0-uLCDM,2017-Nicola-Amara-Refregier-Concordance-quantified} and references therein.

In a recent study \cite{2017-Lin-Ishak-IOI-A}, we reviewed some of the published tensions between data sets and discussed some of the measures of discordance used in the literature. We showed that only comparison between full parameter spaces can measure correctly the degree of discordance between data sets.
We introduced a quantity called the IOI to measure such discordance between two or more data sets and discussed what criteria such a measure must satisfy. We applied there the two-experiment IOI to investigate the inconsistency between geometry and growth of large-scale-structure data sets and found that a moderate to strong inconsistency is present, in agreement with previous works \cite{2015Ruiz-etal-param-splitting,2016Bernal-etal-param-splitting,2013CFHTlens,2017-KiDS-Weak-lensing}.

\begin{table*}[htbp!]
        \caption[Jeffreys's scales for IOI]{\label{table-Jeffrey-scale-IOI}Jeffreys's scales to interpret the values of IOI. We note that we have used an overall moderate  terminology similar to that of Ref.\,\cite{Trotta2008-model-select} and weaker than that used Refs.\, \cite{2013-tension-Verde-etal,2007Heavens-model-selection-forecasting,2016-Hee-etal-model-select}  and the original Jeffreys's scales \cite{jeffreys1998theory}. However, we want to note that ``moderate'' here does not mean ``insignificant'' or ``ignorable,'' but  some tension is present and needs to be resolved. We give below the comparison between IOI and the commonly used confidence level of tension for one-dimensional Gaussian distributions. Note that such a comparison is valid only in one dimension.}
\begin{ruledtabular}
\begin{tabular}{b{0.18\textwidth}|b{0.15\textwidth}b{0.15\textwidth}b{0.15\textwidth}b{0.15\textwidth}}
Ranges &IOI$<1$ & $1<$IOI$<2.5$ & $2.5<$IOI$<5$ & IOI$>5$ \\ \hline
Interpretation &No significant\newline inconsistency&Weak\newline inconsistency &Moderate inconsistency&Strong\newline inconsistency\\
\hline
Confidence level (only) in one dimension & $<1.4$-$\sigma$ & $1.4$-$\sigma-2.2$-$\sigma$ & $2.2$-$\sigma-3.2$-$\sigma$ & $>3.2$-$\sigma$
\end{tabular}
\end{ruledtabular}
\end{table*}

In this paper, we apply the framework developed in \cite{2017-Lin-Ishak-IOI-A} to current data sets in order to have a closer look at some persistent tensions in the literature. We investigate the concordance between and within various groups of data sets of Planck-CMB, large-scale structure and measurements of the Hubble constant. To do so, we follow an algorithmic procedure using IOI measures capable of delineating some possible causes of such inconsistencies.

\begin{figure*}[hbpt!]
\includegraphics[width=0.495\textwidth]{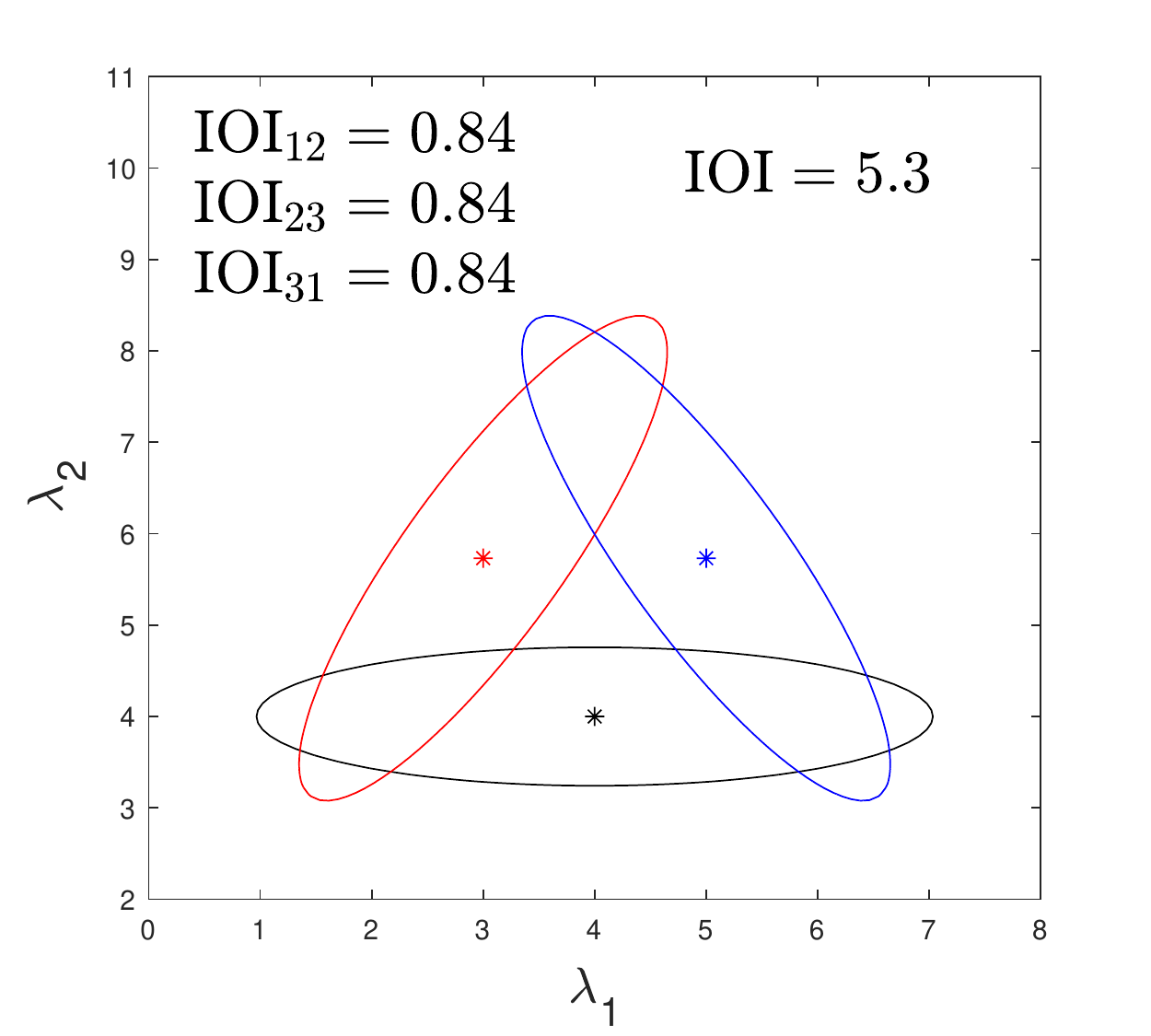}
\includegraphics[width=0.495\textwidth]{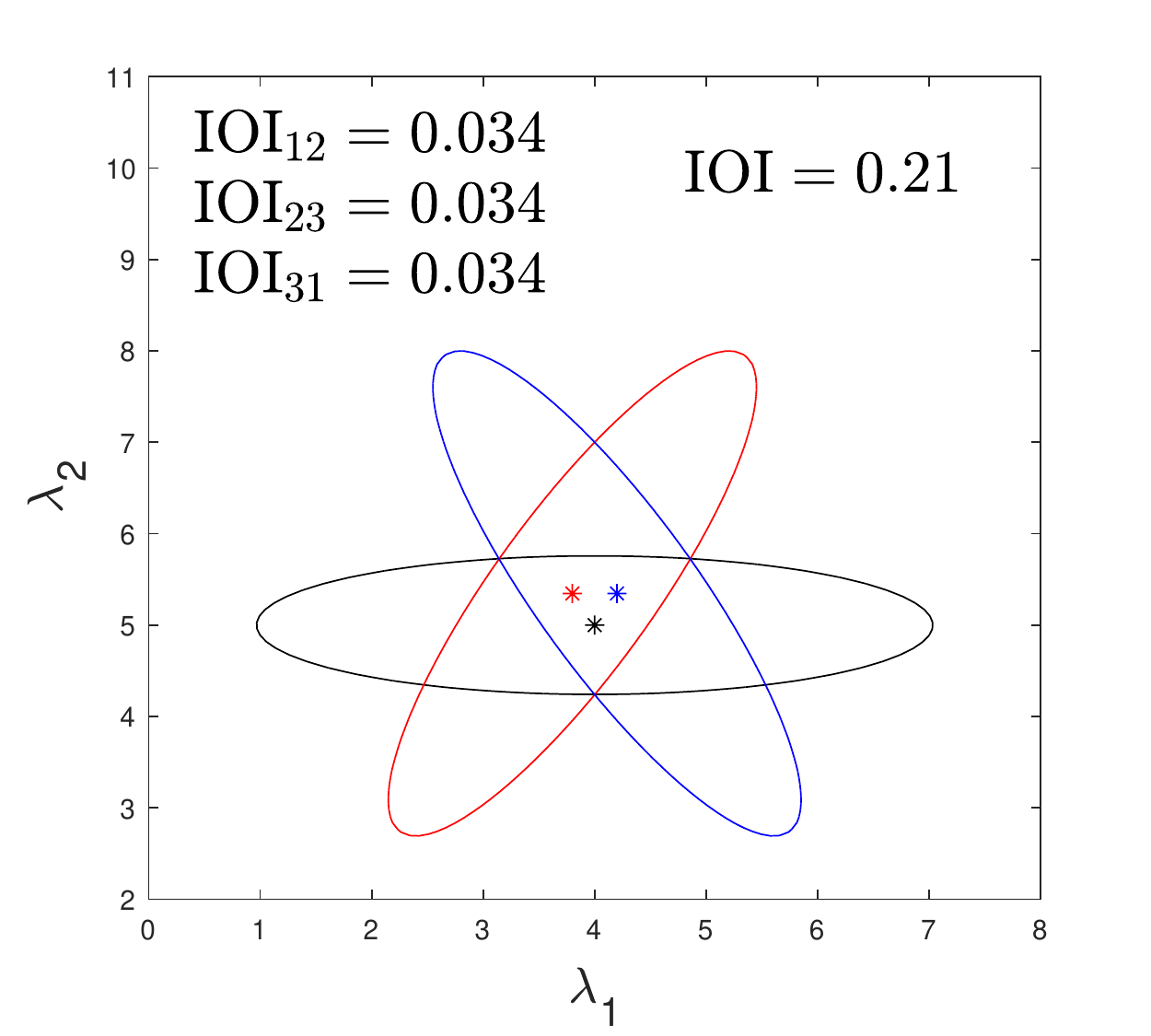}
\caption{\label{fig-multi-IOI-examples}Examples of multiexperiment IOI. As shown above on the left panel, each of the three data sets is consistent with any other one, which is also shown by the three small two-experiment IOIs. But the three data sets are actually inconsistent as a whole because there is little common region simultaneously favored by the three data sets. On the other hand, the three data sets on the right panel are consistent one with another and also consistent when considered as a whole. The three-experiment IOI is able to distinguish between the two situations; IOI=5.3 on the left while IOI=0.21 on the right.}
\end{figure*}

We organize our paper as follows: In Sec.\,\ref{section-methodology}, we briefly review our recently proposed IOI, and discuss how to use it to help identify the cause of inconsistency if presence. In Sec.\,\ref{section-comparison-Hubble}, we investigate the cause of the well-known tension in $H_0$ by comparing five different results, and show mild evidence for a systematic-based origin. In Sec.\,\ref{section-Planck-internal}, we explore the internal inconsistency within the Planck temperature and polarization data sets. In Sec.\,\ref{section-LSS-inconsistency}, discuss an show the consistency between all current LSS data sets. In Sec.\,\ref{section-planck-vs-LSS}, we examine the inconsistency between Planck and each LSS data sets. Based on the moderate inconsistency between Planck and LSS data sets, In Sec.\,\ref{section-tension-forecast}, we perform a forecast of inconsistency between Planck and LSST shear observation. Finally, we give a summary in Sec.\,\ref{section-IOI2-Summary}.

\section{Methodology}\label{section-methodology}
 Our methodology in this work is to compare systematically constraints obtained from various data sets using an algorithmic approach where we employ the IOI measure and other derived quantities. We calculate the IOI between each data sets and the full IOI for the ensemble of the data sets or particular combinations of interest. For various data sets and an underlying model, one can make the following first assertions: (1) if only one or two data sets, while the other data sets are consistent one with another, then it is an indication that such one or two data sets are outliers and (2) if all or most of the data sets are inconsistent with each other, it is an indication that some issues with the underlying model are present.

\begin{figure*}[tp!]
\includegraphics[width=0.99\textwidth]{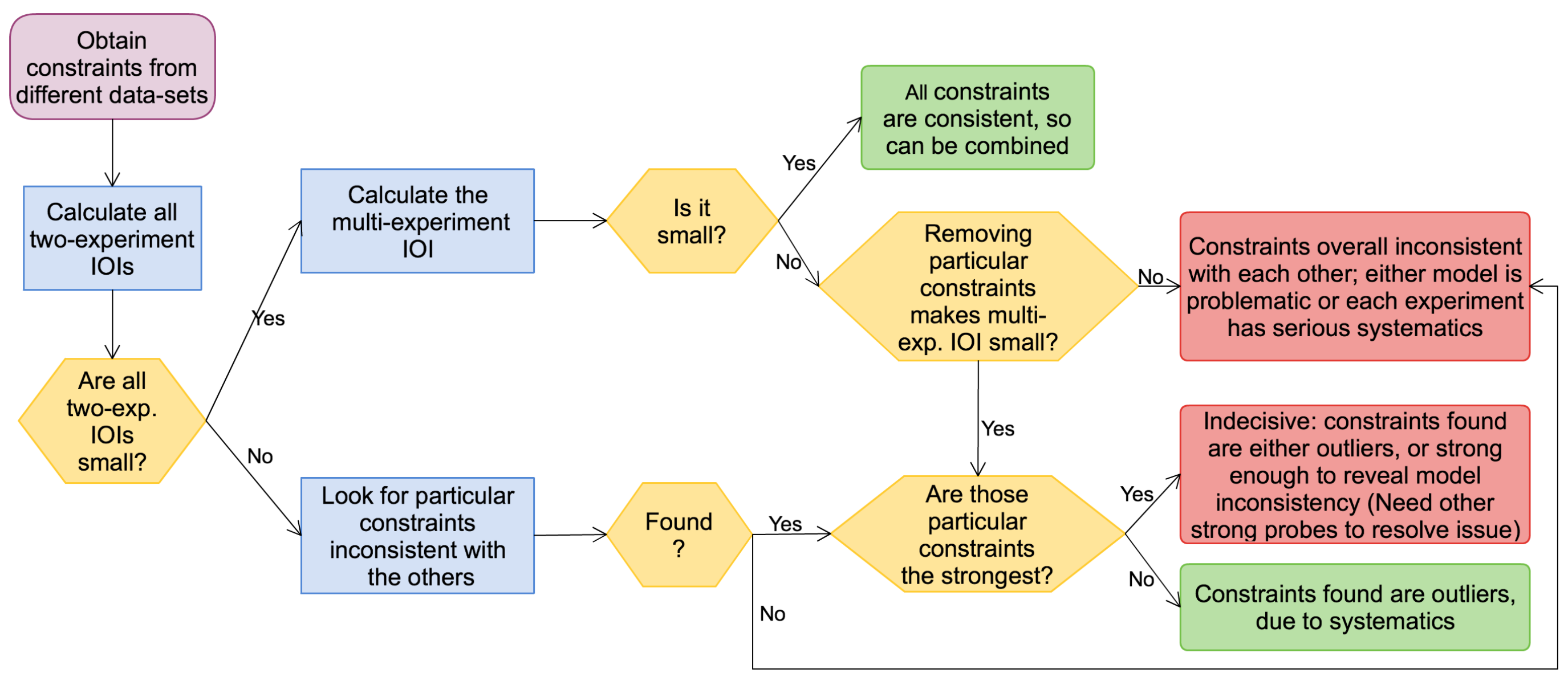}
\caption{\label{fig-IOI_flowchart}A flowchart illustrating the algorithmic method we follow using the IOI measures to help identify the cause of inconsistency, if any. Explanations are provided in the text.}
\end{figure*}

However, the real situation is more complicated and other steps are needed. For example, there are actually two situations in case (1) above, as we explain. If data set A is in tension with others while the others are consistent with each other, it is possible that (1a) data set A is an outlier and (1b) the constraining power of data set A is much stronger than the others. Indeed, it is possible that the underlying model has problems, but only the constraints from data set A are strong enough to reveal the inconsistency due to the underlying model. So if data set A does not fit with other data sets but provides the strongest constraints, we would not have a decisive conclusion. Fortunately, if constraints from a data set are comparable or weaker than others and they are less consistent with others, we can conclude that this data set is an outlier. Finding a cosmological outlier in this way indicate that it suffers from systematic effects that need to be isolated and mitigated. For case (2) above, there are also two possibilities: (2a) the underlying model is problematic and (2b) all data sets have their own systematics causing them to be mutually significantly inconsistent with each other. Fortunately, if we have a large ensemble of data sets to work with, case (2a) is more likely than case (2b). While the proposed method comes with certain limitations, it will become more useful and stringent with future more precise data sets.

\begin{table*}[htpb!]
\caption{\label{table-all-experiments-and-priors}List of data sets considered in this work. Probes are added, where necessary, to break parameter degeneracies.}
\begin{ruledtabular}
\begin{tabular}{llll}
Data sets & \multicolumn{1}{c}{Description} & Added probes\\
\hline
\multicolumn{3}{l}{1. Background}\\
\hline
~~~Local $H_0$ & The locally measured Hubble constant\,\cite{2016Riess-etal-Hubble} & N/A\\
~~~H0LiCOW & Strong gravitational lensing\,\cite{2017-Bonvin-et-al-H0LiCOW} & N/A\\
~~~SN & Supernovae Type Ia \cite{2014supernova740} & BBN\footnote{Primordial deuterium abundance $D/H=2.547\pm0.033\times10^{-5}$ from Ref.\,\cite{2016-Cooke-etal-D-abundance} and helium abundance $Y_p=0.2446\pm0.0029$ from Ref.\,\cite{2016-Peimbert-etal-He-abundance}.} \\
~~~BAO\footnote{We do not use the other two BAO data sets at $z_{eff}=0.32$ and $0.57$ provided by SDSS DR 12 \cite{2016-Gil-Marin-SDSS-DR12} as a background probes. We use them as one LSS probe, SDSS RSD, because BAO measurements are correlated with $f\sigma_8$.} & \multicolumn{1}{p{0.6\textwidth}}{Six Degree Field Glactic Survey (6dF) $(z_{eff}=0.106)$ \cite{2011BAO-6df}\newline and SDSS main galaxy sample (MGS) $(z_{eff}=0.15)$ \cite{2015BAO-sdss-mgs} }& BBN\\
\hline
\multicolumn{3}{l}{2. Planck temperature and polarization}\\
\hline
~~~TT+lowTEB & Planck high-$\ell$\footnote{High-$\ell$ means $30\leq\ell\leq2508$.} temperature auto correlation\,\cite{Planck2015XIII-Cos.Param.} & lowTEB\footnote{The Planck low-$\ell$ ($2\leq\ell\leq29$) temperature and polarization \cite{Planck2015XIII-Cos.Param.}.}\\
~~~TE+lowTEB & Planck high-$\ell$ temperature-E polarization cross correlation\,\cite{Planck2015XIII-Cos.Param.} & lowTEB\\
~~~EE+lowTEB & Planck high-$\ell$ E-mode polarization auto correlation\,\cite{Planck2015XIII-Cos.Param.} & lowTEB\\
~~~TTTEEE+lowTEB & Planck high-$\ell$ temperature and E-mode polarization joint data set \cite{Planck2015XIII-Cos.Param.} & lowTEB\\
\hline
\multicolumn{3}{l}{3. Large scale structure}\\
\hline
~~~WiggleZ   & Power spectrum from WiggleZ Dark Energy Survey \cite{2010WiggleZ-MPK,2012WiggleZ-MPK} & SN+BBN+lowTEB\\
~~~SDSS RSD  & Sloan Digital Sky Survey DR 12 CMASS and LOWZ catalogs \cite{2016-Gil-Marin-SDSS-DR12} & SN+BBN+lowTEB\\
~~~CFHTlens  & CFHTlens survey of cosmic shear/weak lensing \cite{2013CFHTlens} & SN+BBN+lowTEB\\
~~~CMB lens  & Planck 2015 CMB lensing \cite{2016-Planck-galaxy-lensing} & SN+BBN+lowTEB\\
~~~SZ\footnote{When we use the above added probes (SN+BBN+lowTEB) to break dengeneracies in the SZ data set, we do not include priors on $n_s$,  $\Omega_bh^2$ or the cluster mass bias. The cluster mass bias is will be marginalized over. In the last line of the table where we use the joint LSS probes including SZ, priors of $n_s=0.9624\pm0.014$ and $\Omega_bh^2=0.022\pm0.002$ are used and the mass prior from weighting-the-giant \cite{2015-Mantz-etal-SZ}  is applied.}    & Cluster number count from Planck 2015 Sunyaev-Zel'dovich effect \cite{2015Planck-SZ-cluster-count} &  SN+BBN+lowTEB\\
~~~Join LSS & WiggleZ+SDSS RSD+CFHTlens+CMB lens+SZ &   lowTEB or with priors
\end{tabular}
\end{ruledtabular}
\end{table*}

In Ref.\,\cite{2017-Lin-Ishak-IOI-A}, we defined a moment-based new quantity called IOI to measure inconsistencies between two data sets as well as between multiple data sets. The two-experiment IOI is defined as
\begin{equation}\label{eq-two-exp-IOI-definition}
{\rm IOI}=\tfrac{1}{2}\bm{\delta}^T(\bm{C^{(1)}}+\bm{C^{(2)}})^{-1}\bm{\delta}\,,
\end{equation}
where $\bm{C^{(i)}}$ is the covariance matrix of parameters given by the $i$th data set, and $\bm{\delta}$ is the difference of the two parameter means. The multiexperiment IOI is defined as
\begin{equation}\label{eq-multi-exp-IOI-definition}
{\rm IOI}=\frac{1}{N}\Big(\sum\limits_{i=1}^{N}\bm{\mu^{(i)}}\,^T\bm{L^{(i)}}\bm{\mu^{(i)}}-\bm{\mu}^T \bm{L}\bm{\mu}\Big)\,,
\end{equation}
where $\bm{L^{(i)}}=(\bm{C^{(i)}})^{-1}$, and $\bm{L}=\sum\bm{L^{(i)}}$.

In Ref.\,\cite{2017-Lin-Ishak-IOI-A}, we proposed a moderate version of Jeffreys's scales  shown in Table \ref{table-Jeffrey-scale-IOI} as guidelines for the interpretation of values of IOI. Jeffreys's scales were empirical scales originally used as suggestive interpretation of Bayesian evidence ratio \cite{jeffreys1998theory}. We also use these scales as suggestive interpretation of values of IOI because (1) they seem to match well the inconsistencies shown in our sample plots in Ref.\,\cite{2017-Lin-Ishak-IOI-A}; (2) some other measures of inconsistency that reduce or relate to IOI in Gaussian cases also use Jeffreys's scales \cite{2017-Lin-Ishak-IOI-A}, so it is useful to use the same scales for comparison. We note that we use an overall weaker terminology than in the original Jeffreys's scales \cite{jeffreys1998theory} or some other works \cite{2013-tension-Verde-etal,2007Heavens-model-selection-forecasting,2016-Hee-etal-model-select}. We adopt an overall moderate terminology similar the one used in, for example, Ref.\,\cite{Trotta2008-model-select} which is shown in Table \ref{table-Jeffrey-scale-IOI}. But we note that ``moderate'' in Table \ref{table-Jeffrey-scale-IOI} does not mean ``insignificant'' or ``ignorable''. Additionally, we provide for two one-dimensional Gaussian distributions, a one-to-one relation between IOI and the commonly used confidence level of tension, i.e.,
\begin{equation}
\nsigma=\sqrt{2 \,\,{\rm IOI}}.
\label{IOIsigma}
\end{equation}
So in one dimension, a moderate inconsistency means a tension at the 2.2-$\sigma$\ to 3.2-$\sigma$ confidence level. A moderate inconsistency is something we should pay attention to, in particular when it is at the high end of this range.

Given several cosmological data sets, our first step is to calculate the two-experiment IOIs between every two of them. If the two-experiment IOI between data set A and data set B is small, we can conclude that data set A is consistent with data set B. From all two-experiment IOIs, we will be able to see if any particular data set is inconsistent with the others, or if none of them are consistent with any others.

But even if all two-experiment IOIs are small and every two data sets are consistent, we still cannot conclude that those data sets are consistent as a whole, especially in a multiparameter space, as we explain and see also Fig.\,\ref{fig-multi-IOI-examples} for an illustration. On the left panel of Fig.\,\ref{fig-multi-IOI-examples}, any two of the three data sets are consistent with each other, but the three data sets are actually inconsistent when we consider them as a whole. On the other hand, on the right panel of Fig.\,\ref{fig-multi-IOI-examples}, any two of the three data sets are consistent with each other, as well as the three data sets as a whole.  Using the two-experiment IOI alone cannot distinguish the two situations. But multiexperiment IOI defined in Eq.\,\eqref{eq-multi-exp-IOI-definition} can help us tell the difference: the multiexperiment IOI for the left panel is larger than the right. Therefore, besides calculating all the two-experiment IOIs, we also need to calculate the multiexperiment IOI. If the multiexperiment IOI of group X is small, we can conclude that those data sets in that group are consistent with each other.

As an overall summary for the various cases, we show in Fig.\,\ref{fig-IOI_flowchart} an algorithmic flowchart about how to use IOI to help identify the cause of cosmological discordance.

A final comments is that IOI is a moment-based quantity, and requires the distributions on model parameters to be Gaussian or nearly Gaussian. Likely, cosmological data sets usually cannot provide Gaussian distributions alone due to parameter degeneracy that exists with each data set. In order to obtain nearly Gaussian distributions from every data set, it is then necessary to include some other probes in each data set to break the degeneracies between parameters.  The added probes should be (1) a minimum set, so that they are just enough to break the degeneracy in each data set, without strongly influencing the constraints provided by that data set, so that if an inconsistency if found then it should not be due to the added probes; (2) a similar set through all data sets, so that the found inconsistency is not due to the different added probes. For example, for Planck high-$\ell$ temperature and polarization data, we are adding the Planck low-$\ell$ temperature and polarization data (lowTEB).  To keep the consistency of the added probes, we also add to LSS data sets the lowTEB. But adding lowTEB is not enough to break the degeneracies in each LSS data set; therefore, we additionally include with each LSS data set the supernova (SN) and primordial element abundance (BBN) data sets. We will show later that those two additional probes to LSS data sets don't cause any extra inconsistencies between LSS data sets and Planck. We list in Table \ref{table-all-experiments-and-priors} all the cosmological data sets along with the added probes considered in this work.

\section{Hubble constant from five methods}\label{section-comparison-Hubble}
There has been a persistent tension between the local measurement of $H_0$ (local $H_0$ hereafter) and the one derived from Planck constraints with $\Lambda$CDM model \cite{2017Casertano-Riess-Lattanzi-H0-Gaia,2016Riess-etal-Hubble,2016-Bernal-Verde-Riess-H0,2013-tension-Verde-etal}. In order to check whether this inconsistency is due to any statistics in either of the two results, we additionally obtained constraints on $H_0$ from three different methods. The first method is by combining type 1a supernova (SN), baryon acoustic oscillation (BAO) and primordial element abundances from big bang nuclear synthesis(BBN) data. The second one is by combining some large-scale-structure (joint LSS) data sets. And the last one is the Hubble constant measured from gravitational strong lensing time delay.  All methods  assume a $\Lambda$CDM model.

\begin{description}
\item[Planck] The tightest constraint on $H_0$ is derived from Planck cosmic microwave background (CMB) observations assuming the $\Lambda$CDM model. Ref. \cite{Planck2016-intermediate-results-XLVI} provided $H_0=66.93\pm0.62$ km/sec/Mpc using high-$\ell$ temperature and polarization data sets and an updated analysis on Planck low-$\ell$ CMB polarization data.  Note that, in this section, we simply call this constraint Planck. But in other sections, we will explicitly indicate whether Planck temperature or polarization data sets are being used.

\begin{figure}[tpb!]
\includegraphics[width=0.49\textwidth]{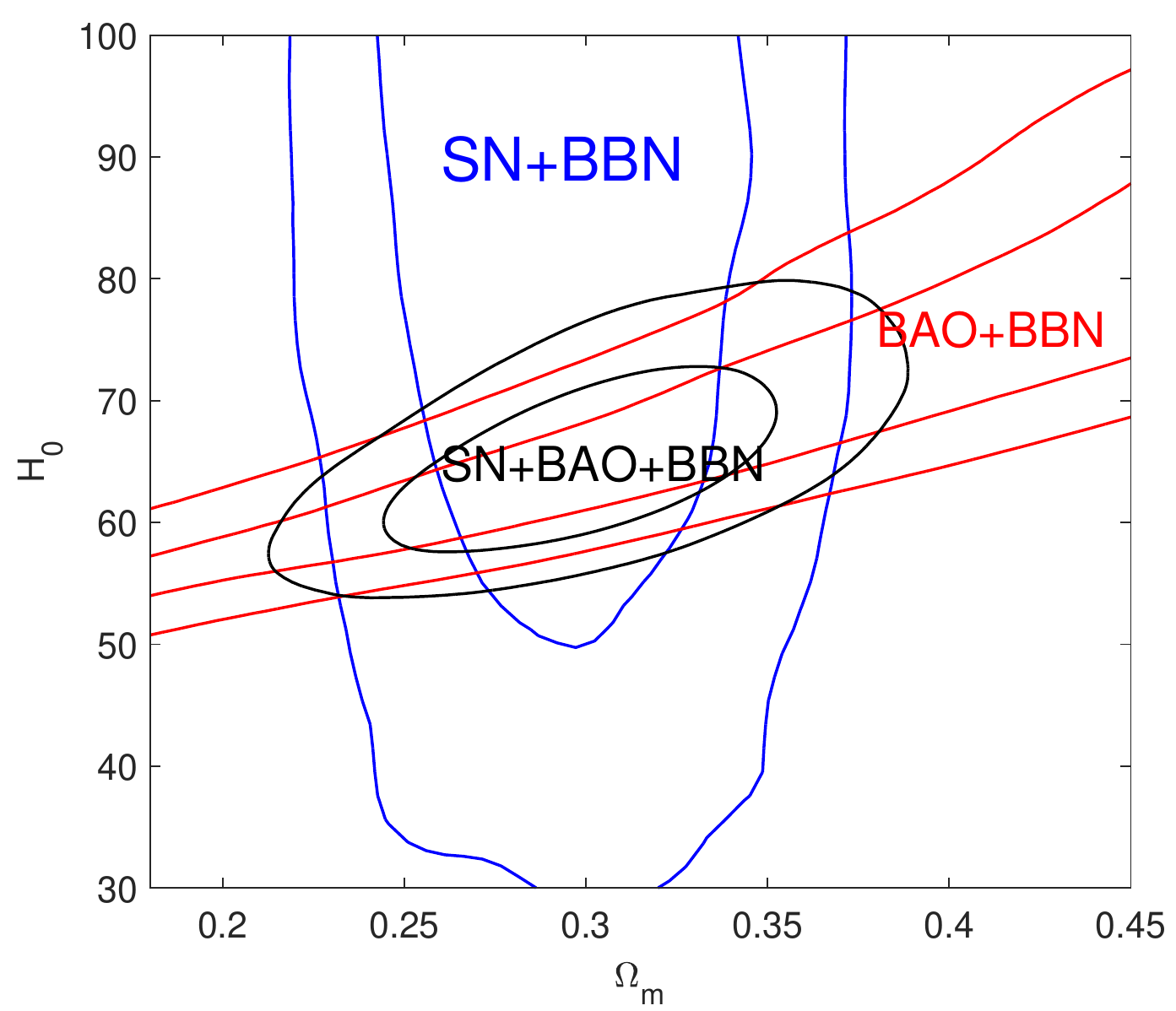}
\caption{\label{fig-SN-vs-BAO}Constraints on $H_0$ vs $\Omega_m$ plane marginalized over $\Omega_bh^2$. Combing SN and BBN (blue) can constrain $\Omega_m$, but not $H_0$. For the constraint from combining BAO and BBN (red), $H_0$ and $\Omega_m$ are positively degenerate. Combing SN, BAO and BBN can then set constraint on $\Omega_m$ and $H_0$, as the blue and red contours break the degeneracies of each other.}
\end{figure}

\item[SN+BAO+BBN] BBN can provide a constraint on $\Omega_bh^2$. Given a value of $\Omega_bh^2$, the constraints from SN and BAO have different degeneracy directions; see Fig.\,\ref{fig-SN-vs-BAO}. So combing SN, BAO and BBN data can break the degeneracy in $\Lambda$CDM model at the backgroud level, without calibrating the distance to supernova or a prior knowledge of the drag sound horizon scale. The value of Hubble constant obtained from this method is $H_0=65.6\pm5.2$ km/sec/Mpc. As a check, even if we fix $\Omega_bh^2=0.02225$ (Planck TTTEEE+lowTEB from Ref.\,\cite{Planck2015XIII-Cos.Param.}) instead of using BBN to constraint it, we still obtain almost the same result $H_0=65.3\pm5.2$ km/sec/Mpc. This is because the correlation between $\Omega_bh^2$ and $H_0$ is small (i.e. $\rho{(\Omega_bh^2,H_0)}=0.06$).

\item[Joint LSS] Combining all the LSS data sets (with prior) is enough to break the degeneracy in each data set alone. We shall see that all the LSS data sets are so far quite consistent with each other (see Sec.\,\ref{section-LSS-inconsistency}), so we can jointly analyze them. The Hubble constant obtained from this method is $H_0=66.48\pm3.99$ km/sec/Mpc.

\item[Local $H_0$] The Hubble constant can be locally measured by ladder distance observation. Different results have been reported, and all have tension with the Planck result to different extents. References.\,\cite{2016Riess-etal-Hubble,2011Riess-etal-hubble} reported that $H_0=73.24\pm1.74~\rm{km~s^{-1}/Mpc^{-1}}$, which is higher than the Planck result at the 3.4-$\sigma$ confidence level.
Ref.\,\cite{2014Efstathiou-Hubble} used a different outlier rejection criteria in the Cepheid samples and found that the high value of locally measured Hubble constant in Ref.\,\cite{2011Riess-etal-hubble} may be due to a systematic error in the calibration. But in Ref.\,\cite{2016Riess-etal-Hubble}, the authors used more Cepheid variables calibrated type Ia supernova and confirmed their earlier high value of Ref.\,\cite{2011Riess-etal-hubble}. Also, Ref.\,\cite{2017Casertano-Riess-Lattanzi-H0-Gaia} used Gaia Data Release 1 and reported a similar high value of the Hubble constant ($H_0=73.0~\rm{km~s^{-1}/Mpc^{-1}}$), different from the Planck measurement at $2.5-3.5$ $\sigma$ level. In this work, we use $H_0=73.24\pm1.74~\rm{km~s^{-1}/Mpc^{-1}}$ as reported in Ref.\,\cite{2016Riess-etal-Hubble}.

\item[Gravitational time delay] The Hubble constant can be measured based on the joint analysis of multiple-imaged quasar systems with measured gravitational time delays. For the $\Lambda$CDM model, Ref.\,\cite{2017-Bonvin-et-al-H0LiCOW} found $H_0=71.9^{+2.4}_{-3.0}$ km/sec/Mpc (see, $H_0$ Lenses in COSMOGRAIL's Wellspring (H0LiCOW) \cite{2017-Bonvin-et-al-H0LiCOW}). This result is more consistent with the local measurement than the derived result from Planck. They also found that fixing $\Omega_m=0.32$ from the Planck result makes their constraint on $H_0$ even higher, $H_0=72.8\pm2.4$ km/sec/Mpc. In this work we the result without fixing $\Omega_m$. But since IOI is a moment-based quantity, we take the average of the upper and the lower uncertainties and use $H_0=71.9\pm2.7$ km/sec/Mpc as the constraints from H0LiCOW.
\end{description}

\begin{table*}[htpb!]
\caption{\label{table-summary-H0}A summary of the constraints on $H_0$ obtained from five different methods. For H0LiCOW, we take the average of the upper and lower uncertainties reported in Ref.\,\cite{2017-Bonvin-et-al-H0LiCOW} and use $H_0=71.9\pm2.7$ km/s/Mpc. Considering the nonaveraged uncertainties on $H_0$ from H0LiCOW would only strengthen our conclusion; see the text for discussion.}
\begin{ruledtabular}
\begin{tabular}{lccccc}
Methods & Planck & SN+BAO+BBN & Joint LSS (with priors) & Local $H_0$ & H0LiCOW\\
\hline
$H_0$ ($\frac{\rm km/sec}{\rm Mpc}$)  & $66.93\pm0.62$& $65.6\pm5.2$ & $67.94\pm1.64$ & $73.24\pm1.74$ & $71.9\pm2.7$
\end{tabular}
\end{ruledtabular}
\caption{\label{table-IOI-H0-mutual}The two-experiment IOIs for $H_0$ obtained from five different methods. There is a strong inconsistency between Planck and local measurement of $H_0$, and a weak inconsistency between Planck and H0LiCOW. The constraint from SB+SN+BBN is relatively wide and consistent with Planck, local $H_0$ and H0LiCOW, but are more consistent with the Planck results than with the two others. The constraint from Joint LSS is consistent with Planck and H0LiCOW, but show some small inconsistency with local $H_0$. Planck has much stronger constraint on $H_0$ rendering it more demanding for (in)consistency tests than any other probes, but it is  still provides a constraint on $H_0$ that is more consistent with SB+SN+BBN and LSS  compared to the local measurement of $H_0$.}
\begin{ruledtabular}
\begin{tabular}{l|rrrrr}
  & Planck & SB+SN+BBN & Joint LSS & Local $H_0$ & H0LiCOW\\
  \hline
Planck & --- & 0.03   &    0.17    &    5.83  &  1.61 \\
SB+SN+BBN & 0.03   & ---   &   0.09    &    0.98  &  0.58 \\
Joint LSS & 0.17 &  0.09 & ---   &  2.46  &  0.79\\
Local $H_0$ & 5.83   &   0.98    &  2.46& --- & 0.09\\
H0LiCOW & 1.61  &  0.58  &  0.79 &  0.09  & ---
\end{tabular}
\end{ruledtabular}
\caption{\label{table-IOI-H0-multi}Multiexperiment IOIs for constraints on $H_0$ obtained from five different methods. The multiexperiment inconsistency for all data sets is mainly caused by the inconsistency between Planck and local $H_0$. This is reflected by the fact that IOI drops after removing either the local measurement of $H_0$ or Planck, and that IOI increases after removing any of the other three data sets. 
We find that removing local $H_0$ leads to the smallest multiexperiment IOI. And since the constraint from the local measurement on $H_0$ is not the strongest, one can conclude that the local measurement of $H_0$ is likely to be an outlier as discussed in the text.}
\begin{ruledtabular}
\begin{tabular}{l|cccccc}
Data sets & ~All & \multicolumn{1}{b{0.1\textwidth}}{Removing\newline Planck}& \multicolumn{1}{b{0.13\textwidth}}{Removing\newline SB+SN+BBN}& \multicolumn{1}{b{0.1\textwidth}}{Removing\newline LSS}& \multicolumn{1}{b{0.1\textwidth}}{Removing\newline Local $H_0$}& \multicolumn{1}{b{0.1\textwidth}}{Removing\newline H0LiCOW}\\
\hline
Multiexperiment IOI & 2.85 & 1.52 & 3.51  & 3.56 & 0.88 & 2.96 
\end{tabular}
\end{ruledtabular}
\end{table*}

The constraints on $H_0$ from the above five different methods are summarized in Table \ref{table-summary-H0}. We can see that the results obtained from SN+BAO+BBN and joint LSS are closer to the one obtained from Planck than to the local measurement of $H_0$ or the ones from HOLiCOW. We also calculate the two-experiment IOIs and multiexperiment IOIs for the five data sets, which are shown in Table \ref{table-IOI-H0-mutual} and Table \ref{table-IOI-H0-multi}.

From the two-experiment IOIs shown in Table \ref{table-IOI-H0-mutual}, we can see that there is a strong inconsistency (IOI=5.83) between the Planck result and the local $H_0$ and a weak inconsistency between Planck and H0LiCOW. Using the relation between IOI in one dimension and the commonly used confidence level of tension, i.e. Eq.\,\eqref{IOIsigma}, the strong inconsistency IOI=5.83 corresponds to a tension at the 3.4-$\sigma$ confidence level which has been reported in Ref.\,\cite{2016Riess-etal-Hubble}. For the other two methods SN+BAO+BBN and joint LSS, due to their weaker (large) constraints on $H_0$, their results are overall consistent with Planck, the local $H_0$ and H0LiCOW. However, they are more consistent with the Planck result than the local $H_0$ or H0LiCOW as reflected on the various values of IOIs.

The inconsistency within the five results of $H_0$ is also shown by the multiexperiment IOI shown in Table \ref{table-IOI-H0-multi}. The multiexperiment IOI for the five results is 2.86. This inconsistency is mainly caused by the inconsistency between the Planck and local measurements of $H_0$. That is because if we remove either Planck or the local $H_0$ and recalculate the multiexperiment IOI for the remaining four results, we get a much lower IOI (i.e. IOI=1.52 after removing Planck and IOI=0.85 after removing the local $H_0$). Also if we remove any one of the other three results, IOI rather increases. So it tends to suggest that either Planck or local $H_0$ measurement is an outlier; see the further discussion below. 

Our analysis favors that the local measurement of $H_0$ is an outlier. This follows our reasoning described in Sec.\,\ref{section-methodology}, and the facts that the multiexperiment IOI drops the most after removing the local measurement of $H_0$, and that the constraint from the local measurement is not the strongest. Therefore, there is an indication that the local $H_0$ is an outlier. Removing Planck also drops the multiexperiment IOI, but not as much as the case of removing the local measurement. Also Planck's constraint is the strongest among all the five results, it renders Planck the most demanding one in (in)consistency tests. Is it possible that there is problem with the underlying model? Probably not, because removing Planck would have led to the lowest multiexperiment IOI if it were the case.

We note that for H0LiCOW we have taken the average of its upper and lower constraint limits of $H_0$ since IOI is a moment-based quantity. But we consider the actual constraint from H0LiCOW, our conclusion will only be strengthened. That is because the actual constraints from H0LiCOW has a larger lower uncertainty than upper uncertainty. So H0LiCOW is actually slightly more consistent with Planck and less consistent with local $H_0$ than what is shown in our analysis. This would support further that the local measurement of $H_0$ is an outlier.

There are studies suggesting that some extension of the $\Lambda$CDM model can resolve the tension on $H_0$ between Planck and the local measurement; see for examples Refs.\, \cite{2016-Hubble-reconcile,2016-Valentino-Melchiorri-Silk-reconciling-Planck-local,2017-Yang-Pan-Mota}. Our analysis, however, shows that the local measurement of $H_0$ is likely an outlier, favoring a systematic-based explanation for such a tension. Our result is consistent with the analysis by the authors in Refs.\,\cite{2001-Gott-Vogeley-Podariu-Ratra,2003-Chen-Gott-Ratra,2011-Chen-Ratra,2017-Chen-Kumar-Ratra}. The authors in Ref.\,\cite{2011-Chen-Ratra} used median statistics on 553 measurements of $H_0$ and found $H_0=68\pm5.5$ km/sec/Mpc at the 95\% statistical and systematic errors. The authors in Ref.\,\cite{2017-Chen-Kumar-Ratra} used 28 Hubble parameters at different redshifts ($0.07\leq z\leq2.3$) and found that $H_0=68.3^{+2.7}_{-2.6}$ km/sec/Mpc in the $\Lambda$CDM model. The results of those analyses are more

It is interesting that despite moderate inconsistencies between Planck and joint LSS (to be explored in Sec.\,\ref{section-planck-vs-LSS}), their constraints on $H_0$ are found consistent here, IOI=0.17. But as we explained in Ref.\,\cite{2017-Lin-Ishak-IOI-A} that for a multidimensional model, even if we see consistency on a parameter in a marginalized space, there can still be inconsistency due to that parameter. We introduced in Ref.\,\cite{2017-Lin-Ishak-IOI-A} two single-parameter measures: the residual IOI$_i$ and the drop $\Delta$IOI$_i$. If two data sets have no inconsistency due to one particular parameter, both measures should be small. That small value of 0.17 is actually IOI$_{H_0}$. And we found that $\Delta$IOI$_{H_0}=0.07$, which is also small. Since both single-parameter IOIs (here IOI$_{H_0}$ and $\Delta$IOI$_{H_0}$) are small, it is justified to conclude that Planck and joint LSS are consistent on $H_0$, even though there are inconsistencies between them in the full parameter space. Those inconsistencies reside in other parameters.

It is worth pointing out that previous studies have shown that the $w$CDM model can resolve the tension  on $H_0$ between Planck and the local measurement; see Refs.\,\cite{2017-Zhao-etal-nature-astronomy,2017-Valentino-etal-DEdynamics-constraints,2016-Valentino-Melchiorri-Silk-reconciling-Planck-local} and also Sec.XI.B in our previous work Ref.\,\cite{2017-Lin-Ishak-IOI-A}. However as we pointed out in Ref.\,\cite{2017-Lin-Ishak-IOI-A} that such a lower tension in the $w$CDM model is mainly caused by degraded uncertainty on $H_0$. 

In sum, we conclude, in this section, that by comparing the above constraints on $H_0$ obtained from five different methods, there is some indication that the constraints on $H_0$ is an outlier.

\section{Internal (in)consistency between Planck temperature and polarization data sets}\label{section-Planck-internal}
The CMB temperature and polarization observations from Planck provide powerful constraints on cosmological parameters. Reference \cite{2016-Shafieloo-Hazra-Consistency-Planck} used a technique called cross statistics to investigate the internal (in)consistency between Planck CMB temperature and polarization data sets. They found that the temperature best fit in $\Lambda$CDM is consistent with the EE+lowTEB data, while EE+lowTEB best fit is not consistent with that of the temperature data. They pointed out that this is due to the large uncertainty in the polarization data. In this section, we will use IOI to examine the level of (in)consistency between the Planck temperature and polarization data sets.

For CMB temperature and polarization observations, Planck provides three high-$\ell$ correlation data sets: temperature auto correlation (high-$\ell$ TT), temperature and E-mode polarization cross correlation (high-$\ell$ TE), and E-mode polarization auto correlation (high-$\ell$ EE). Since the temperature data set cannot constrain the reionization optical depth, we include in those three correlation data sets the low-$\ell$ temperature and polarization data (lowTEB). We provide here the two-experiment IOIs for the three data sets, along with their multiexperiment IOI. We calculate IOIs with with the standard $\Lambda$CDM cosmological parameters: $\Omega_bh^2$, $\Omega_ch^2$, $\theta$, $\log_{10}(A_s)$, $n_s$ and $\tau$.

From Table \ref{table-IOI-Planck-internal}, we can see that there is a moderate inconsistency between TT+lowTEB and EE+lowTEB data sets. The TE+lowTEB data set is more consistent with the TT+lowTEB data set than with the EE+lowTEB data data set. But this is probably because of the fact that the TT+lowTEB data set has stronger constraints than the EE+lowTEB data set, so that the TE+lowTEB cross correlation is closer to the TT+lowTEB data set. The multiexperiment IOI is 3.33, which again tells us there would be a moderate inconsistency among the three data sets.

\begin{table}[tpb!]
\caption{\label{table-IOI-Planck-internal} IOIs between Planck temperature and polarization data sets. We can see that the Planck EE+lowTEB data set has a moderate inconsistency with the Planck TT+lowTEB data set, IOI=2.61. The multiexperiment IOI for the Planck TT+lowTEB, TE+lowTEB and EE+lowTEB is even higher, IOI=3.34.}
\begin{ruledtabular}
\begin{tabular}{l|ccc}
Planck & TT+lowTEB & TE+lowTEB  & EE+lowTEB\\
 \hline
 TT+lowTEB  & --- & 0.94 & 2.61  \\
 TE+lowTEB  & 0.94  & --- & 1.93  \\
 EE+lowTEB   & 2.61 & 1.93  & --- \\
 \hline
 \multicolumn{4}{l}{Multiexperiment IOI = 3.34}
\end{tabular}
\end{ruledtabular}
\end{table}

For the high-$\ell$ EE data set, we do not need to include lowTEB to break the degeneracy. If we constrain the six $\Lambda$CDM parameters using high-$\ell$ EE data set without lowTEB, the IOI between it and TE+lowTEB with lowTEB becomes slightly larger, IOI = 3.38. This slight increased IOI is mainly caused by the increased difference in the constraints on $\tau$.

\begin{table*}[tpb!]
\caption{\label{table-IOI-LSS} Two-experiment IOIs for LSS data sets. We recall that we also add to each LSS data set the lowTEB+SN+BBN to break degeneracies; see Table \ref{table-all-experiments-and-priors}. All IOIs are below unity (i.e. not significant on Jeffreys's scales), which means any two of the current LSS data sets are consistent one with another.}
\begin{ruledtabular}
\begin{tabular}{l|ccccc}
 & WiggleZ & SDSS RSD & CFHTlens & CMB lens & SZ \\
 \hline
WiggleZ  & ---       &  0.25    &  0.60    &  0.49   &  0.33  \\
SDSS RSD &  0.25     &  ---     &  0.21    &  0.92   &  0.45  \\
CFHTlens &  0.60     &  0.21    &  ---     &  0.69   &  0.66  \\
CMBlens  &  0.49     &  0.92    &   0.69   &  ---    &  0.42 \\
SZ       &  0.33     &  0.45    &   0.66   &   0.42  &  ---
\end{tabular}
\end{ruledtabular}
\caption{\label{table-IOI-LSS-multi} Multiexperiment IOIs for LSS data sets. The multiexperiment IOI for the five LSS data sets shows again the consistencies between the LSS data sets. Considering the results from this table and those from the two-experiment IOIs shown in Table \ref{table-IOI-LSS}, we conclude that the LSS data sets are all consistent one with another and also consistent when considered as a whole.}
\begin{ruledtabular}
\begin{tabular}{l|cccccc}
Data sets & All  & \multicolumn{1}{b{0.12\textwidth}}{Removing\newline WiggleZ} & \multicolumn{1}{b{0.1\textwidth}}{Removing\newline SDSS RSD} & \multicolumn{1}{b{0.1\textwidth}}{Removing\newline CFHTlens} & \multicolumn{1}{b{0.1\textwidth}}{Removing\newline CMB lens} & \multicolumn{1}{b{0.09\textwidth}}{Removing\newline SZ}\\
\hline
Multiexperiment IOI & 0.96  & 0.99  & 0.83 & 0.87  & 0.54 & 1.00
\end{tabular}
\end{ruledtabular}
\end{table*}

Recently, authors in Ref.\,\cite{2016-Addison-etal} pointed out that there is tension within the TT data set between the higher multiple power spectrum ($1000\leq\ell\leq2508$) and the lower multiple  power spectrum ($2\leq\ell<1000$). They also show that a lensing anomaly parameter $A_{\rm lens}$ greater than unity can resolve such a tension. It is possible that the tension between the higher and the lower TT multiple power spectra is responsible for the internal moderate inconsistency within Planck both temperature and polarization data set found in this work. Further investigations will be devoted to find the real causes of the internal inconsistencies within Planck.

We conclude in this this section that there is a moderate inconsistency between Planck temperature (TT) and polarization (EE) data. But this does not tell us which data set is (or both are) biased by systematics. This agrees with what was found in Ref.\,\cite{2016-Shafieloo-Hazra-Consistency-Planck} that there is a mild amplitude difference between Planck temperature and polarization data sets. Our conclusion about the Planck internal inconsistency, which is based on the full parameter space, is however slightly stronger than that in Ref.\,\cite{2016-Shafieloo-Hazra-Consistency-Planck}.

\section{Consistency between Large-scale-structure data sets}\label{section-LSS-inconsistency}
In this section, we investigate the consistencies among LSS data sets. Exploring the mutual inconsistencies among LSS data sets is an important step before we use them jointly to constrain cosmological parameters. In Table \ref{table-all-experiments-and-priors}, we list the five probes categorized as LSS data sets including: the Power spectrum from WiggleZ Dark Energy Survey (WiggleZ) \cite{2010WiggleZ-MPK,2012WiggleZ-MPK}, Sloan Digital Sky Survey data release 12 CMASS and LOWZ catalogs (SDSS RSD) \cite{2016-Gil-Marin-SDSS-DR12}, CFHTlens survey of cosmic shear/weak lensing with the most conservative priors (CFHTlens) \cite{2013CFHTlens}, Planck 2015 CMB lensing \cite{2016-Planck-galaxy-lensing} and Cluster number count from Planck 2015 Sunyaev-Zel'dovich effect \cite{2015Planck-SZ-cluster-count}. Differently from Planck CMB temperature and polarization observations, LSS data sets probe the late-time structure growth. For these reasons, we categorize CMB lens and SZ into LSS data sets.

We show the two-experiment IOIs between every two LSS data sets in Table \ref{table-IOI-LSS}. We can see that every two LSS data sets are consistent one with another as  all the two-experiment IOIs are below unity (no significant inconsistency on Jeffreys's scales). But as we discussed in Sec.\,\ref{section-methodology}, for multiparameter model, even if every two data sets are consistent one with another, there is still possible inconsistencies when we consider all the data sets together as a whole. We therefore calculate the multiexperiment IOIs and show them in Table \ref{table-IOI-LSS-multi}. The multiexperiment IOI for all LSS data sets is also found to be below unity. This again leads us the conclusion that current LSS data sets are all  consistent.

In Sec.\,\ref{section-comparison-Hubble} we have already used the joint constraint on $H_0$ from LSS data sets. But it is actually the consistency among current LSS data sets found in this section that allows us to safely set joint constraints on cosmological parameters from various LSS data sets.  Finally, to explore further, we successively remove one LSS probe at a time and recalculate the multiexperiment IOI as shown in Table \ref{table-IOI-LSS-multi}. We do not find any significant drop of multiexperiment IOI by removing any of LSS probes. This means again no particular LSS probe is significantly inconsistent with the rest.

We conclude, in this section, that the current LSS data sets are all consistent in constraining cosmological parameters.

\section{Planck vs Large-scale-structure data sets}\label{section-planck-vs-LSS}
Several inconsistencies between Planck and LSS data sets have been pointed out in the literature; see, for examples, Refs.\,\cite{2013CFHTlens,2017-KiDS-Weak-lensing,2011WiggleZ-growth-rate,2016Bernal-etal-param-splitting, 2014sdssIII-redshift-space,2015Planck-SZ-cluster-count}. In this section, we use IOI measures to explore these  inconsistencies between Planck and individual LSS data set as well as the joint LSS data sets.

\begin{table*}[tpb!]
\caption{\label{table-IOI-CMB-vs-LSS} Two-experiment IOIs for Planck temperature and polarization vs LSS data sets. Compared to Table \ref{table-IOI-LSS}, we can see that the inconsistencies between Planck data sets and LSS data sets are generally larger than those between the LSS data set themselves. We also note that in order to break the degeneracy in each LSS data sets, we include in each of them the probes lowTEB+SN+BBN. }
\begin{ruledtabular}
\begin{tabular}{l|ccccc}
\multicolumn{6}{l}{Planck vs LSS+lowTEB+SN+BBN  }\\
  \hline
         & WiggleZ & SDSS RSD & CFHTlens & CMB lens  & CMB SZ\\
\hline
TT+lowTEB       & 3.09  &  3.72   & 3.53  &  2.92  & 3.07  \\
TE+lowTEB       & 1.38  &  1.87   & 1.57  &  0.99  & 1.34  \\
EE+lowTEB       & 2.01  &  2.30   & 2.03  &  1.51  & 1.95  \\
TTTEEE+lowTEB   & 3.44  &  4.20   & 4.05  &  3.76  & 3.46
\end{tabular}
\end{ruledtabular}
\begin{ruledtabular}
\caption{\label{table-IOI-CMB-vs-LSS-fixbh2} Similar to Table \ref{table-IOI-CMB-vs-LSS}, but here $\Omega_bh^2$ is fixed to 0.02222. We do so to see how the small inconsistency between Planck and BBN is affecting the inconsistencies found between Planck and LSS data sets. IOIs here are calculated in a five-parameter space without $\Omega_bh^2$.  Compared to Table \ref{table-IOI-CMB-vs-LSS}, inconsistencies of similar level still persist between the Planck and LSS data sets even if slightly smaller after fixing $\Omega_bh^2$.}
\begin{tabular}{l|ccccc}
\multicolumn{6}{l}{Planck vs LSS+lowTEB+SN+(fixed $\Omega_bh^2=0.02222$)  }\\
\hline
         & WiggleZ & SDSS RSD & CFHTlens & CMB lens  & CMB SZ\\
\hline
TT+lowTEB       & 2.59  & 3.10  & 3.53  & 2.50  & 2.51\\
TE+lowTEB       & 1.15  & 1.62  & 1.67  & 0.91  & 1.18\\
EE+lowTEB       & 1.64  & 2.14  & 1.19  & 0.66  & 1.18\\
TTTEEE+lowTEB   & 2.87  & 3.50  & 4.00  & 3.27  & 2.81
\end{tabular}
\end{ruledtabular}
\begin{ruledtabular}
\caption{\label{table-IOI-Planck-CMB-lens-Alens}Two-experiment IOIs between Planck (temperature and polarization) and CMB lens (+SN+BBN+lowTEB). Different from the [CMB lens] column in Table \ref{table-IOI-CMB-vs-LSS}, here the constraints from CMB lens are obtained by additionally varying $A_{\rm lens}$ besides the six standard $\Lambda$CDM parameters. The two-experiment IOIs here are larger to those of the [CMB lens] column in Table \ref{table-IOI-CMB-vs-LSS}. This means varying $A_{\rm lens}$ in the CMB lens data set increases the inconsistency between Planck (temperature and polarization) and CMB lens.}
\begin{tabular}{l|cccc}
 & TT+lowTEB & TE+lowTEB & EE+lowTEB & TTTEEE+lowTEB \\
 \hline
CMB lens & 3.37 & 1.56 & 2.13 & 3.83
\end{tabular}
\end{ruledtabular}
\caption{\label{table-IOI-Planck-BAO-lss-Alens} Two-experiment IOIs between Planck temperature and polarization+lowTEB+BAO and LSS data sets calculated in the $\Lambda$CDM+$A_{\rm lens}$ model. Compared to the last row in Table \ref{table-IOI-CMB-vs-LSS}, IOIs are reduced. Compared to Table \ref{table-Planck-vs-joint-LSS} below, the IOI between Planck and joint LSS (with lowTEB) is also lowered. However, weak to moderate inconsistencies still remain.}
\begin{ruledtabular}
\begin{tabular}{l|cccccc}
 & WiggleZ & SDSS RSD & CFHTlens & CMB lens  & CMB SZ & joint LSS (with lowTEB)\\
 \hline
Planck+BAO & 2.38        &        2.65        &        2.72        &        2.11      &       2.45  &  3.04
\end{tabular}
\end{ruledtabular}
\caption{\label{table-Planck-vs-joint-LSS} Two-experiment IOIs between the joint LSS (with lowTEB or with priors) and Planck (temperature and polarization). In the case of joint LSS+lowTEB, the five LSS data sets are combined with lowTEB added to break degeneracies and to constrain $\tau$. A moderate inconsistency (high end of moderate on Jeffreys's scales) exists between joint LSS (with lowTEB) and Planck TTTEEE+lowTEB. In the case of joint LSS+priors, the lowTEB is not used, but $\tau$ is fixed to 0.078 and priors of $\Omega_bh^2=0.022\pm0.002$ and $n_s=0.9624\pm0.014$ are used. Then the inconsistency between joint LSS (with priors) and Planck TTTEEE+lowTEB becomes smaller but still in the moderate range on Jeffreys's scale.}
\begin{ruledtabular}
\begin{tabular}{l|cccc}
 & TT+lowTEB & TE+lowTEB & EE+lowTEB & TTTEEE+lowTEB\\
 \hline
Joint LSS (with lowTEB) & 3.85 & 1.58 & 1.13 & 4.81 \\
Joint LSS (with priors) & 1.93  & 0.80  & 0.64  & 2.83
\end{tabular}
\end{ruledtabular}
\end{table*}

We show the results of two-experiment IOIs in Table \ref{table-IOI-CMB-vs-LSS}. 
The inconsistencies between Planck TT+lowTEB and each LSS data set are moderate to the high end of moderate with IOI ranging from 2.92 to 3.72 on Jeffreys's scales. For the TE+lowTEB and EE+lowTEB data sets, IOIs between them and LSS data sets are smaller than that for TT. Such smaller inconsistencies may be only due to the weaker constraints from TE+lowTEB and EE+lowTEB data sets.
The inconsistencies between Planck (TT, TE or EE) and each LSS data set are larger than those between any two LSS data sets. However, this does not necessarily mean Planck (TT, TE or EE) data sets are outliers since they provides the strongest constraints on cosmological parameters, which makes them most demanding in (in)consistency tests. If we combine the Planck temperature and polarization data sets (TTTEEE+lowTEB) and compare individual LSS probes then IOI increases to 3.44-4.20.

One could question if part of the weak to moderate inconsistencies between Planck (temperature and polarization) and each LSS data set may partially be caused by the inconsistencies between Planck and the added probes (lowTEB, SN and BBN) to each LSS data set. In particular, the constraint on $\Omega_bh^2$ from BBN is somehow inconsistent with that from Planck TT+lowTEB or TTTEEE+lowTEB. Indeed, the constraint on $\Omega_bh^2$ from BBN is $0.0226\pm0.0004$, which is different from $0.02222\pm0.00023$ obtained from Planck high-$\ell$+lowTEB at about 1-$\sigma$ confidence level. To address this, we additionally do the analysis with $\Omega_bh^2$ fixed to 0.02222 in each LSS data set instead of using BBN to constrain it. Then we calculate again the two-experiment IOIs in a parameter space without $\Omega_bh^2$, and show them in Table \ref{table-IOI-CMB-vs-LSS-fixbh2}, by fixing $\Omega_bh^2$ for LSS data sets, we obtain only slightly smaller IOIs for most of LSS data sets but the moderate inconsistencies still persist. Therefore adding BBN to each LSS does not significantly affect the inconsistency between Planck and each LSS.

In an earlier study, authors in Ref.\,\cite{2008-lensing-anomaly} found a lensing anomaly in WMAP data set with a parameter $A_{\rm lens}=3.1^{+1.8}_{-1.5}$ at the 2-$\sigma$ confidence level. But in the recent analysis of Planck 2015 data sets, this anomaly reduced to $A_{\rm lens}=1.22\pm0.1$ at 1-$\sigma$ confidente level \cite{Planck2015XIII-Cos.Param.}. We have seen in Table \ref{table-IOI-CMB-vs-LSS} that there are moderate inconsistencies between Planck TT+lowTEB or TTTEE+lowTEB data sets and Planck CMB lensing data set. In this work, we investigate in particular whether varying the parameter $A_{\rm lens}$ in the analysis of CMB lens data set is able to reduce such inconsistencies. We first obtain constraints on the $\Lambda$CDM + $A_{\rm lens}$ model from CMB lens (+SN+BBN+lowTEB), and then calculate the IOI between Planck (temperature and polarization) and CMB lens with the standard six $\Lambda$CDM parameters. We list the results in Table \ref{table-IOI-Planck-CMB-lens-Alens}. Releasing $A_{\rm lens}$ makes the constraints on the standard six $\Lambda$CDM parameters weaker due to the degeneracies between $A_{\rm lens}$ and other parameters. This should make the IOI smaller. However, the IOIs in Table \ref{table-IOI-Planck-CMB-lens-Alens} are generally larger than those in the ``CMB lens'' column in Table \ref{table-IOI-CMB-vs-LSS}. The reason is that varying $A_{\rm lens}$ in the analysis of CMB lens (+SN+BBN+lowTEB) makes the means of the six $\Lambda$CDM parameters located slightly further from those obtained from Planck (temperature and polarization). Therefore, the constraints from CMB lens (+SN+BBN+lowTEB) with a varying $A_{\rm lens}$ are slightly more inconsistent with those from Planck (temperature and polarization). Reference \cite{Planck2015XIII-Cos.Param.} showed that $A_{\rm lens}$ is slightly above unity when constrained jointly by Planck temperature and polarization along with CMB lensing data sets. One might think that varying $A_{\rm}$ should have reduces the tension between Planck temperature and polarization data sets. This is not necessarily the case, because it is possible that the moderate inconsistency between Planck (temperature and polarization) and CMB lens is what makes $A_{\rm lens}$ greater than unity when the two data sets are jointly analized. Also it is shown in Ref.\,\cite{2016-Addison-etal} that, although a nonstandard value of $A_{\rm lens}\simeq1.4$ can resolve the tension between the higher- ($1000\leq\ell\leq2058$) and the lower- ($2\leq\ell\leq1000$) multiple of power spectra of Planck temperature data, it can not resolve the tension between Planck temperature and CMB lens data sets.

Authors in Ref.\,\cite{2015-Valentino-Melchorri-Silk-Beyon-LCDM} studied cosmological constraints in an extended cold dark matter ($e$CDM) model for a number of data set combinations. Their $e$CDM model consists of 12 cosmological parameters including $A_{\rm lens}$. They pointed out that the value of $\sigma_8$ can be lowered in this $e$CDM model as constrained by, e.g., Planck (temperature and polarization)+BAO, which could potentially resolve the tensions between Planck and LSS data sets. While we reproduced their constraints on the 12 cosmological parameters with Planck+BAO, we would like to know whether varying only $A_{\rm lens}$ in addition can resolve the tensions between Planck and LSS data sets. To do so, we first obtain the constraints on the $\Lambda$CDM+$A_{\rm lens}$ model from Planck TTTEEE+lowTEB+BAO (Planck+BAO). We then calculate the IOIs between Planck+BAO and LSS data sets in the five-parameter space ($\Lambda$CDM without ). We show the results in Table \ref{table-IOI-Planck-BAO-lss-Alens}. Compared to the last row in Table \ref{table-IOI-CMB-vs-LSS} (also Table \ref{table-Planck-vs-joint-LSS} to be described in the next paragraph), IOIs are now lowered. But weak to moderate inconsistencies still remain.

Now let us investigate whether Planck (TT, TE and EE) data sets are consistent with the joint LSS data sets. In Sec.\,\ref{section-LSS-inconsistency} we have shown that the current LSS data sets are consistent with each other, so that we can combine them to jointly constrain cosmological parameters. By combining all LSS data sets we do not need SN or BBN, but lowTEB is still needed to break degeneracies and to constrain $\tau$. We combine the five LSS data sets (joint LSS) with lowTEB added to constrain the six cosmological parameters, and then calculate the IOI between Planck (TT, TE and EE) and joint LSS (lowTEB). We list the IOIs between Planck and joint LSS  (with lowTEB) in Table \ref{table-Planck-vs-joint-LSS}. The IOI between Planck TT+lowTEB and joint LSS (with lowTEB) is 3.85, which is in the high end of moderate inconsistency. The joint LSS (with lowTEB) is more consistent with Planck TE+lowTEB (IOI=1.58) and EE+lowTEB (IOI=1.13) than with Planck TT. Bu this is probably due to their weaker constraints. The joint Planck TTTEEE+lowTEB is more inconsistent with the joint LSS (IOI=4.81) than TT+lowTEB alone. Taking this 4.81 at face value, we have an inconsistency at the very high end of the moderate range on Jefferey's scale.

Alternatively, for the joint LSS, we can fix $\tau=0.078$ without using lowTEB and include priors of $\Omega_bh^2=0.022\pm0.002$ and $n_s=0.9624\pm0.014$ (which can be included in the SZ likelihood). In this case, we also include the weighting-the-giant mass bias \cite{2015-Mantz-etal-SZ} in the SZ likelihood.  We name this combination as joint LSS (with priors).  Without using lowTEB, the constraints (now in a five-parameter space) from joint LSS become weaker. We then expect the inconsistencies between Planck and joint LSS (with priors) become smaller. We also list the IOIs between Planck and joint LSS (with priors) in Table \ref{table-Planck-vs-joint-LSS}, and they are indeed so. All the IOIs with Planck (temperature or polarization) for joint LSS (with priors) are larger than those for joint LSS (with lowTEB). For example, the IOI between Planck TTTEEE+lowTEB and joint LSS (with priors) is 2.83, which is a moderate inconsistency on Jeffreys's scale. So even, without using lowTEB (but with priors in $\Omega_bh^2$ and $n_s$), an moderate inconsistency between Planck TTTEEE+lowTEB and joint LSS exists.

It is worth to point out that an updated mean value of $\tau=0.058$ is reported in Ref.\,\cite{Planck2016-reionization}. We however use the earlier mean value of $\tau=0.078$ reported in Ref\,\cite{Planck2015XIII-Cos.Param.} for two reasons as follows. First, our likelihood analysis on the Planck data set is the same as Ref\,\cite{Planck2015XIII-Cos.Param.}, so we use $\tau=0.078$ to match the constraint obtained from the Planck data set here. Second, the choice of the $\tau$ value would not significantly affect the constraint obtained from joint LSS (with priors), because it can only affect the likelihood of CMB lens among the five LSS probes. So, we expect that using $\tau=0.058$ would only slightly change the results.

Recently, the authors of Ref.\,\cite{2017-Charnock-Battye-Moss} have investigated the (in)consistency between Planck and LSS data sets. They worked with two sets of LSS combinations calling one as \textit{strong set} and the second as \textit{weak set}. Mainly, the strong set contained CFHTLenS (with strong prior on astronomical uncertainties \cite{2013CFHTlens}), Planck CMB lensing, SDSS RSD (DR12) and SZ galaxy cluster counts with the mass bias from Planck CMB lensing \cite{2016-Planck-galaxy-lensing}. The weak set contained the same data sets as the strong one but with the most conservative assumptions in CFHTlens \cite{2013CFHTlens} (that is also what we have used in this analysis) and the weighting-the-giant mass bias \cite{2015-Mantz-etal-SZ} instead of Planck lensing. They used various measures and found, for example, that the \textit{tension}, i.e. $\log T$, which is equivalent to our IOI in the Gaussian limit, indicates a strong tension ($\log T=7.56$) for the strong joint LSS data set and moderate tension ($\log T=2.59$) for the LSS  weak set. 
We comments on the similarities and differences between our work and Ref.\,\cite{2017-Charnock-Battye-Moss} in three aspects. 
(1) Our joint LSS (with priors) data sets combination is similar with their weak joint LSS case. The only difference is that we also included WiggleZ in our joint LSS. Our IOI is in this case 2.83 between Planck TTTEEE+lowTEB and joint LSS (with priors), which is more consistent with their $\log T=2.59$ result. Our joint LSS (with lowTEB) combination additionally has lowTEB added to break degeneracies and to constrain $\tau$. Our IOI between Planck TTTEEE+lowTEB and joint LSS (with lowTEB) lays between their weak and strong results.
(2) Our method and steps in this work are different from Ref.\,\cite{2017-Charnock-Battye-Moss}. Beside testing the (in)consistency between different groups of data sets, we use an algorithmic procedure to search for outliers among these groups. So before combining LSS data sets, we first investigate their inconsistency one with another. After confirming the consistency between all LSS data sets, we then compare the joint LSS to Planck. 
(3) Although Ref.\,\cite{2017-Charnock-Battye-Moss} applied different concordance measures, they seemed to focus their discussion mainly on the \textit{difference vector} qunatitity introduced in Ref.\,\cite{2015-Battye-Charnock-Moss-difference-vector}. But when they used the \textit{tension} ($\log T$) then our results using the IOI measure are found between their two cases with their weak-set results being close to ours considering almost the same data sets. 

We conclude in this section: (1) There are moderate inconsistencies between the Planck CMB baseline data sets and LSS data sets (individual or joint). Including lowTEB in joint LSS data sets to break degeneracies leads to higher inconsistency between Planck and joint LSS data sets. (2) Planck CMB polarization data sets (TE+lowTEB and EE) are more consistent with LSS data sets than the temperature (TT) data set, but probably because of their weaker constraints; (3) If we combine CMB temperature and polarization data sets, the inconsistencies between Planck TTTEEE+lowTEB and LSS data sets (individual or joint) become even stronger than the inconsistencies between Planck TT+lowTEB and LSS data sets (individual or joint). (4) So far, we cannot determine whether Planck CMB observation is an outlier, or the underlying model is problematic, but only Planck's constraints are strong enough to reveal it.

\section{Forecasting the level of tension between Planck and LSST cosmic shear}\label{section-tension-forecast}
In this section, we forecast the level of possible inconsistency between Planck and future cosmic shear observation by the Large Synoptic Survey Telescope (LSST) \cite{LSST-web} and the $\Lambda$CDM as the underlying cosmological model. We use the Fisher matrix formalism to forecast the constraints that will be obtained by LSST cosmic shear observations, taking the fiducial cosmological parameters as the mean values obtained from the current joint LSS (with lowTEB) data sets. We made this choice of fiducial parameter values because current LSS observations (include cosmic shear) are found consistent with each other. Then we estimate the inconsistency between Planck and future LSST cosmic shear by calculating IOI between them. If the best fits of the $\Lambda$CDM parameter from future LSST cosmic shear truly locates at such fiducial values, the inconsistency between Planck and LSST cosmic shear would become significantly strong.

The Fisher matrix for cosmic shear tomography reads \cite{1999-Hu-sh-tomography}
\begin{equation}\label{eq-Fisher-matrix-sh-tomo}
F_{\alpha\beta}=\sum_{\ell}(\ell+1/2)\Delta\ell f_{sky}{\rm tr}[\bm{R}^{-1}\bm{C}^{\rm shear}_{,\alpha}\bm{R}^{-1}\bm{C}^{\rm shear}_{,\beta}]\,
\end{equation}
where $\bm{C}^{\rm shear}(\ell)$ is the tomographic cosmic shear angular power spectra, and $\bm{R}=\bm{C}^{shear}+\bm{N}$. The element of noise power spectra is $N_{ij}=\tfrac{\sigma_{\epsilon}^2}{n_i}\delta_{ij}$, with $\sigma_{\epsilon}$ the rms intrinsic shear and $n_i$ is the source-galaxy number density per steradian in photometric-redshift bin $i$. We choose 12 logarithmically separated $\ell$ bins centered from $\ell=20$ to $\ell=5000$. We divide a range of photometric-redshift from $z_{ph}=0$ to $z_{ph}=3.5$ into five bins, with bin width $\Delta z_{ph}= 0.4$ for the first four bins and $1.6<z_{ph}<3.5$ for the last. The specifications of LSST are listed in Table \ref{table-LSST-specification} \cite{LSST-web}. We do not consider marginalizing over photometric-redsift systematic parameters, or the impact of intrinsic alignment; see Refs.\,\cite{2016-Krause-etal-IA-impact,2017-Yao-Lin-Ishak-Troxel,2015-Troxel-Ishak-lensing,2015-KirK-etal-Galaxy-aligments,TroxelEtAlCross} for discussions on impact of intrinsic alignment. So the estimated constraints for LSST cosmic shear are based on an ideal situation. If photometric-redshift error and intrinsic contamination were considered, the estimated constraints for LSST cosmic shear would be larger (see, for example, Ref.\,\cite{2016-Krause-etal-IA-impact}), and the resulting IOI would be smaller. So our estimates of IOIs are conservative. After calculating the Fisher matrix, the estimated covariance matrix of cosmological parameters is obtained as
\begin{equation}\label{eq-covariance-matrix-inverse-Fisher}
\bm{C}=\bm{F}^{-1}\,.
\end{equation}
\begin{table}[tpb!]
\caption{\label{table-LSST-specification}Specification of LSST; see \cite{LSST-web}.}
\begin{ruledtabular}
\begin{tabular}{l c c c c c}
Specifications & $f_{sky}$ & $\sigma_{\epsilon}$ & $n_{gal}/\rm{arcmin^2}$ & $z_0\footnote{Effective survey depth.}$ & $z_{max}$\\
Values & 0.436 & 0.26 & 26 & 0.5 & 3.5
\end{tabular}
\end{ruledtabular}
\end{table}
We compute the shear power spectra using \textsc{CosmoSIS} \cite{2015-cosmosis}. We fix $\Omega_b$ to 0.0483, since LSST cosmic shear can only poorly constrain it. 
We then use the covariance matrix from Eq.\,\eqref{eq-covariance-matrix-inverse-Fisher} and take the fiducial model parameters as the mean values in order to calculate the IOI between Planck and LSST cosmic shear. We do that in the four-dimensional parameter space: $\Omega_m$, $\sigma_8$, $H_0$ and $n_s$. We find a very high IOI of 17 indicating a very strong inconsistency on Jeffreys's scales.  

As we discussed in Sec.\,\ref{section-planck-vs-LSS}, there is currently an inconsistency on the high-moderate end on Jeffreys's scales between Planck and joint LSS data sets with IOI=4.81. We find here that if the best fit cosmological parameters from LSST cosmic shear are those obtained from the current LSS data sets, then there would be a very strong inconsistency between Planck and LSST cosmic shear with an IOI=17.

\section{Summary}\label{section-IOI2-Summary}

We proposed and used a systematic procedure employing the two- and multiexperiment IOI in order to investigate (in-)consistencies between various data sets of Planck, Large-Scale Structure and the Hubble parameter measurements. The algorithmic method can delineate the cause of the inconsistencies in some cases. We applied the tests within and across the three groups of data sets. The detailed results, discussions and implications are provided in the sections above while we provide here a short summary. 

\begin{enumerate}

\item We compared the constraints on $H_0$ from five different methods. We found a significant drop of the multiexperiment IOI when removing the local measurement of $H_0$. This provides some indication that the local measurement of $H_0$  is an outliers. More explicitly, the multiexperiment IOI drops from 2.85 for all five constraints to 0.88 when the local measurement of $H_0$ is removed .  Thus the tension in $H_0$ between Planck and the local measurement of $H_0$ is likely due to systematics in the local measurement. We however note that this is a moderate indication and that more precise future constraints on $H_0$ should be able to confirm if this is the indeed the case.

\item We found a moderate inconsistency (IOI=2.61) between the temperature (TT) and the polarization (EE) data sets within Planck that needs to be resolved. This is in agreement with Ref.\,\cite{2016-Shafieloo-Hazra-Consistency-Planck} which found a mild amplitude difference comparing temperature and polarization data.

\item We found that the current LSS data sets, such as WiggleZ power spectrum, SDSS redshift space distortion, CFHTLenS weak lensing, CMB lensing and cluster count from Sunyaev-Zel'dovich effect are consistent one with another and also when all combined together. This means we can safely use them to jointly constrain cosmological parameters and rely on their joint results.

\item When the full parameter space in the $\Lambda$CDM model is considered, there are moderate inconsistencies between Planck TT+lowTEB and LSS individual or joint data sets, with IOIs ranging from 2.92 to 3.72. Planck polarization data sets (TE+lowTEB and EE+lowTEB) have smaller inconsistencies with LSS data sets than the TT+lowTEB, because they have weaker constraining powers. Planck joint TTTEEE+lowTEB and individual LSS data sets have IOIs ranging from 3.44 to 4.20.  We found IOI=4.81 between Planck TTTEEE+lowTEB and joint LSS (+lowTEB), which is an inconsistency on the high end of the moderate range (on Jeffreys's scales) and must be resolved with future precise data sets. If priors on $\Omega_m$ and $n_s$ are used with $\tau$ fixed to 0.078 instead of adding lowTEB , the inconsistency between Planck TTTEEE+lowTEB and joint LSS becomes smaller although it persists with an IOI of 2.83.

\item We finished the analysis with a forecast study on how LSST cosmic shear measurement will be able to constrain the degree of possible inconsistencies between large-scale structure (LSS) and Planck. We found that if the best fit of LSST cosmic shear is similar to that of the current joint LSS (+lowTEB) data sets, we would have a very strong inconsistency (IOI=17) between LSST and Planck. Future LSS surveys such as LSST will provide us with highly decisive answers on the level of (dis)concordance between Planck and LSS. Also, future CMB experiments, e.g. \cite{2017-COrE,Stage-IV-paper1,PIXIE2011}, will provide more constraining powers and will have a say into this as well.
\end{enumerate}

Some of these inconsistencies seems to persist. They can be due to systematic effects in the data or to the underlying model. One way or the other, they need to be resolved as we keep moving toward precise and accurate cosmology.

\begin{acknowledgments}
M.I. acknowledges that this material is based upon work supported in part by the National Science Foundation under Grant No. AST-1517768 and an award from the John Templeton Foundation.
\end{acknowledgments}

\bibliography{IOI,IOI_II}{}

\end{document}